\documentclass[sigconf]{acmart}
\usepackage{dcohen}
\usepackage{microtype}
\usepackage{graphicx}
\usepackage{placeins}
\usepackage{amsthm}
\usepackage{subfigure}
\usepackage{float}
\usepackage{multirow}
\usepackage{booktabs} 

\definecolor{mint}{rgb}{0.24, 0.71, 0.54}

\AtBeginDocument{%
  \providecommand\BibTeX{{%
    \normalfont B\kern-0.5em{\scshape i\kern-0.25em b}\kern-0.8em\TeX}}}

\setcopyright{acmcopyright}
\copyrightyear{2021} 
\acmYear{2021} 
\setcopyright{acmcopyright}
\acmConference[SIGIR '21]{Proceedings of the 44th International ACM SIGIR Conference on Research and Development in Information Retrieval}{July 11--15, 2021}{Virtual Event, Canada}
\acmBooktitle{Proceedings of the 44th International ACM SIGIR Conference on Research and Development in Information Retrieval (SIGIR '21), July 11--15, 2021, Virtual Event, Canada}
\acmPrice{15.00}
\acmDOI{10.1145/3404835.3462951}
\acmISBN{978-1-4503-8037-9/21/07}

\newtheorem{thm}{Theorem}[section]

\begin{document}

\title{Not All Relevance Scores are Equal: Efficient Uncertainty and Calibration Modeling for Deep Retrieval Models}

\author{Daniel Cohen}
\affiliation{%
  \institution{Brown University}
  \city{Providence, R.I.}
   \country{USA}}
\email{daniel_cohen@brown.edu}

\author{Bhaskar Mitra}
\affiliation{%
  \institution{Microsoft}
  \city{Montreal}
   \country{Canada}}
\email{bmitra@microsoft.com}

\author{Oleg Lesota}
\affiliation{%
  \institution{Johannes Kepler University}
  \city{Linz}
   \country{Austria}}
\email{oleg.lesota@jku.at}

\author{Navid Rekabsaz}
\affiliation{%
  \institution{Johannes Kepler University}
  \institution{Linz Institute of Technology, AI Lab}
  \city{Linz}
   \country{Austria}}
\email{navid.rekabsaz@jku.at}

\author{Carsten Eickhoff}
\affiliation{%
  \institution{Brown University}
  \city{Providence, R.I.}
   \country{USA}}
\email{carsten@brown.edu}

\renewcommand{\shortauthors}{Cohen et al.}

\begin{abstract}
In any ranking system, the retrieval model outputs a single score for a document based on its belief on how relevant it is to a given search query. While retrieval models have continued to improve with the introduction of increasingly complex architectures, few works have investigated a retrieval model's belief in the score beyond the scope of a single value. We argue that capturing the model's uncertainty with respect to its own scoring of a document is a critical aspect of retrieval that allows for greater use of current models across new document distributions, collections, or even improving effectiveness for down-stream tasks. In this paper, we address this problem via an efficient Bayesian framework for retrieval models which captures the model's belief in the relevance score through a stochastic process while adding only negligible computational overhead. We evaluate this belief via a ranking based calibration metric showing that our approximate Bayesian framework significantly improves a retrieval model's ranking effectiveness through a risk aware reranking as well as its confidence calibration. Lastly, we demonstrate that this additional uncertainty information is actionable and reliable on down-stream tasks represented via cutoff prediction.
\end{abstract}





\maketitle

\section{Introduction}

Recent work in neural information retrieval (IR) models have achieved impressive performance on a variety of retrieval tasks whether the models are based on pre-trained Transformer architectures~\cite{repbertZhan,BERTPassageRankCho,ColBERTKhattab,learning2retrievezhan2020} or learned from scratch~\cite{TransformerKernelDocHofstatter20,convKNRMDai18, mitra2020conformer}. These state-of-the-art models treat their estimates of a document's relevance as a deterministic score. While effective, this deterministic perspective obfuscates a large amount of critical information that a user could use to determine whether their query is effective or when they have gone so far down a ranked list that the model is no longer sure of its scores. Ideally, an effective IR system should be able to gracefully convey when it is no longer effective or confident in its rankings, which is a substantial risk given neural models' vulnerability to out-of-distribution inputs from a different collection or even a new first stage ranker~\cite{BERT_NSkarpukhin20,BayesianLastLayerKristiadi,cohen_adv}.

\begin{figure}
    \centering
    \includegraphics[width=\linewidth]{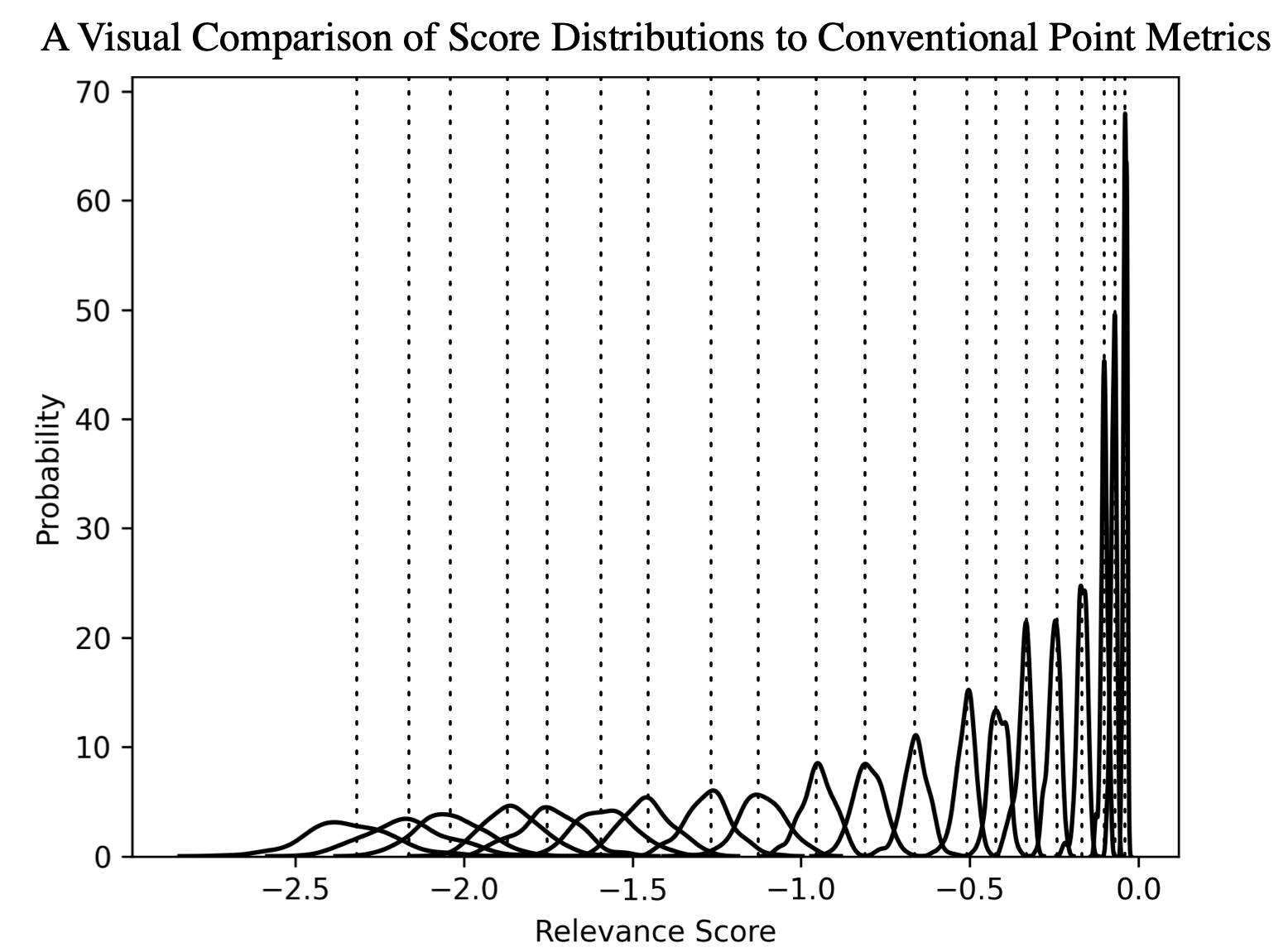}
    \caption{A visual comparison of the conventional interpretation of neural model rankings which provides a single score for each document (represented by the dashed vertical line), and a Bayesian perspective which captures the uncertainty over each score. Highly relevant documents have a narrow score distribution as the retrieval model is confident in itself. In the case of documents where it is uncertain, the model is able to convey its uncertainty by means of a wider distribution of possible relevance scores. The plot represents a single query and select candidate documents from TREC Deep Learning Track 2019 with multiple raw score estimates per document produced by a Bayesian mini BERT.}
    \label{fig:bayesian}
\end{figure}

In order to fulfill the above criteria, a retrieval model should be capable of displaying this \textit{uncertainty} over document relevance prediction through a distribution of possible scores as demonstrated in Figure~\ref{fig:bayesian}.
Furthermore, the retrieval model should be expressive enough such that the mean score of the document should convey the model's belief in the actual relevance while its corresponding variance captures the model's uncertainty -- a high variance should indicate that the model is unsure of a document's relevance even if it is highly placed in a ranked list.
Beyond this expression of uncertainty, calibration is another desirable quality as a comparatively low relevance score with respect to another candidate document should reflect a proportional likelihood that the lower scoring document is less relevant. 

These concepts of calibration and uncertainty have been touched on in previous work in IR, most specifically query performance prediction~\cite{QPPBoostStrapSDRoitman,QPPStandardDevShtok,QPPHamedNeural} and query cutoff prediction~\cite{bicut_cutoffLien19,choppycutoffBahri20,cutoffCulpepperCL16}. In these settings, an external model attempts to capture a portion of the model's uncertainty as a function of the input data and its deterministic output. These post-retrieval approaches rely on scoring a large number of documents while simultaneously only representing uncertainty over the documents rather than the model itself. 

In contrast, ensemble methods such as models featuring on MS MARCO leaderboards\footnote{https://microsoft.github.io/msmarco/} capture both types of uncertainty -- aleatoric which is uncertainty over the input documents as mentioned above and epistemic uncertainty which is uncertainty over the parameters of a model. Unfortunately, ensemble methods become computationally expensive as retrieval models grow in size and complexity, and results in a substantial obstacle given the prevalence of BERT and other large transformer architectures~\cite{BERT_NSkarpukhin20,BERTPassageRankCho,ColBERTKhattab,BERTT5S2SDocNogueira20}. As the objective is to maximize performance while ranking as many top $n$ documents as possible, running $m$ models simultaneously results in $\frac{n}{m}$ fewer ranked documents. Lastly, in cases where these large models are used as pre-trained encoders, ensembles do not adequately capture uncertainty over these shared parameters.

As such, we approach capturing uncertainty from a Bayesian perspective where leveraging a convenient property of dropout~\cite{DropoutHinton14} can be treated as a form of variational inference, referred to as Monte Carlo dropout~\cite{BayesianMCDropoutGal}. In this setting, dropout induces a stochastic ranking model which creates a distribution of scores as the dropout mechanism outputs different values each time it is run over the same input candidate document. The characteristics of this distribution then allow us to capture both aleatoric and epistemic uncertainty. 
While MC sampling still relies on an infeasible number $m$ of forward passes over the ranking model, we expand on this work with a theoretically motivated extension where only the last layer needs to be Bayesian to capture both epistemic and aleatoric uncertainty. By efficiently sampling from the last layers, we are able to not only attain a distribution of scores on par with standard deterministic models, but also leverage this uncertainty information to improve both final rankings and a downstream task of cut off prediction with a risk aware reranking.

Succinctly we summarize our research contributions as:

\begin{enumerate}
\item An efficient approximate Bayesian framework for any neural model trained with pairwise cross entropy or pairwise hinge loss.
\item A rank based calibration metric that facilitates uncertainty comparison across retrieval models.
\item A risk aware reranking method that significantly improves reranking performance.
\item Exposing the actionable information contained in uncertainty for downstream tasks via cut off prediction.
\end{enumerate}

\section{Related Work}

Given the high prevalence of probabilistic ranking approaches in IR, the formal concept of uncertainty for retrieval was first discussed by \citet{RiskAwareIRZhu09}. In their work, they treat the variance of a probabilistic language model~\cite{ponte1998language} as a risk-based factor to improve its performance for retrieval. However, relying on a generative model's self reported uncertainty in modern models often results in high calibration errors~\cite{GenKnowWhatTheyDontKnowNalisnick19}. Furthermore, they assume all document uncertainty to be normal in nature, whereas we demonstrate that a significant distribution shift occurs across rank positions. A modern perspective on this is through exploiting neural generative IR models. While the relevance estimation mechanism in current neural/deep ranking models are rooted in text-to-text matching principles, the probabilistic/generative paradigm views relevance estimation as the probability of generating query given document. This paradigm has a long history in IR research, initiating from \citet{ponte1998language}. In this regard, few recent works have provided a modern interpretation of generative IR models, primarily through exploiting sequence-to-sequence neural generative models. For instance, \citet{dos2020beyond} exploit large-scale sequence-to-sequence Transformer-based models to rank answers according to their generation probability for given a question, while \citet{nogueira2019document} use a sequence-to-sequence model to first generate queries conditioned on a document, and then use them to expand the document. The probabilistic nature of generative models makes them readily capable of estimating the aleatoric uncertainty. A natural way of estimating uncertainty in generative models is by exploiting the entropy of the next term prediction distributions, defined over the vocabulary set, as proposed by \citet{izacard2020leveraging} in the context of abstractive summarization. In the light of studying uncertainty in IR, we also investigate the characteristics of this self-reported entropy-based uncertainty using a BERT-based deep generative IR model for the downstream task of cutoff prediction. Lastly, we highlight the simultaneous work of~\citet{penha2021calibration}, where they explore the impact of Bayesian and ensemble uncertainty in the conversational response space under BERT. In contrast to our work, their approach requires a full forward pass of BERT for every Monte Carlo sample.

In approaches that consider both probabilistic and discriminative models, a close parallel in spirit are the tasks of query performance prediction (QPP) and cutoff prediction. In QPP, the aim is to determine how difficult a query is given a collection. Within QPP, the concept of post-retrieval query prediction, which relies on the output of one or more retrieval models, is closest to our work. A variety of works have investigated this problem; however, no QPP method has directly incorporated a retrieval model's uncertainty.
 
For example, \citet{QPPStdCUmmins11} examine the concept of the distribution of scores by modeling the standard deviation of all candidate document scores to estimate the difficulty of the query. \citet{QPPAslamShannonMultScoring07} use an ensemble of multiple ranked lists to predict the difficulty of a query by examining the diversity through Jensen-Shannon divergence. In subsequent work,~\citet{QPPmeanDocScoreHaggai17} attempt to achieve a mean document score for each query, introducing the notion of calibrated scores for retrieval. An alternative perspective that attempts to capture this uncertainty is via query perturbation approaches~\cite{QPPperturbZhou06}. These methods inject noise into the initially ranked documents to determine the robustness of the ranked list which sheds some light on both aleatoric and epistemic uncertainty.

In a similar vein to QPP, cutoff prediction attempts to identify the optimal cutoff point to maximize some non-monotonic metric such as $F_1$ or to determine a set of candidate documents to pass on to the next stage for cascade based retrieval~\cite{bicut_cutoffLien19,cutoffCulpepperCL16,choppycutoffBahri20}. The motivating hypothesis is that retrieval models become increasingly volatile and unstable as documents drift further from the training distribution. As such, external models are trained to identify this instability and determine when the model is no longer reliable while attempting to learn the retrieval model's uncertainty through deterministic document scores. 

Finally, in the context of classical IR models enhanced with word embeddings, \citet{rekabsaz2017exploration} showcase the benefits of exploiting the uncertainty of word-to-word similarities for identifying a reliable threshold to filtering highly similar terms. In their work, the uncertainty is defined as the variance over the similarities achieved from an ensemble of neural word embeddings, all trained under the same learning configuration.
\vspace{-10pt}
\section{Measuring Retrieval Uncertainty}

In this section, we first define the problem and motivation prior to introducing the efficient Bayesian framework. As this framework relies on specific assumptions about the loss function used to train a retrieval model, we extend this work to not only work on pairwise cross entropy, but pairwise hinge loss to cover a wide spectrum of model architectures.

Subsequently, we discuss the issue of traditional calibration metrics when considering the ranking problem and posit an alternative metric. Finally, we discuss our risk aware reranking method that leverages the uncertainty information produced from the approximate Bayesian framework.

\subsection{Problem Statement and Motivation}

Let $\dcmc{Q} = \{q_1, \ldots, q_n \} $ be the set of queries, and $\dcmc{C} = \{d_1, \ldots, d_m\}$ be the collection of documents some retrieval model parameterized by $\dcvec{\theta} \in\dcbb{R}^G$ is trained to retrieve over. Our dataset is then $\dcmc{D} = \dcmc{Q} \times \dcmc{C}$, such that $f_\theta : \dcmc{Q} \times \dcmc{C} \rightarrow \dcbb{R}$ produces a score for each query-document pair evaluated. The task is then to find an ordering of scores that maximizes an external metric such as user satisfaction or relevance and is approximated via mean average precision, nDCG, or MRR among other functions. The vast majority of retrieval models rely on stochastic weight optimization to achieve a maximum likelihood estimates (MLE) or maximum a posteriori (MAP) approximation for $\dcvec{\theta}$ configuration. While effective, this setting produces point estimates for each candidate in $\dcmc{D}$, removing all uncertainty and confidence estimates, from the predictions. At this point, areas of research such as QPP and cut off prediction try to determine these quantities heuristically via score and document distributions~\cite{cutoffCulpepperCL16,bicut_cutoffLien19,QPPAslamShannonMultScoring07,QPPBoostStrapSDRoitman,QPPHamedNeural,QPPStandardDevShtok,QPPNeuralnfQAHashemi}. This task has become increasingly challenging with the changing nature of neural retrieval models as previously established post-retrieval QPP methods are not as effective for neural models~\cite{QPPHelia}.

We therefore propose capturing this uncertainty information directly from the model by enforcing a Bayesian view on a portion of $\dcmc{\theta}$. In doing so, the distribution over our weights induces a distribution over our candidate scores, allowing for downstream uncertainty and confidence estimates for each document. The remainder of this section covers our efficient Bayesian retrieval framework, uncertainty calibration algorithm, and risk aware reranking.

\subsection{Efficient Bayesian Neural Retrieval}

We first introduce Bayesian inference, and then an efficient interpretation that allows for measuring uncertainty in real world environments. As discussed, the conventional process of training a retrieval model $f_\theta$ is through a form of stochastic gradient descent to achieve an estimate of the MLE over some dataset $\mathcal{D}$, $P(\dcmc{D}|\mathbf{\theta}),$ or MAP if regularized, $P(\dcmc{D}|\mathbf{\theta})P(\mathbf{\theta}),$ which minimizes some loss function,
$$\mathbf{\theta} = \argmin_\theta \dcmc{L}(\dcmc{D}, f, \mathbf{\theta}).$$\\
This representation of $f_\theta$ unfortunately discards all other parametrizations of $\dcvec{\theta}$ which are potentially just as capable of determining the relevance of a document. The hypothesis is that some parameterizations will be better for scoring some subset of query-document pairs while other parameterizations will excel on other areas of the dataset. The disparity between scores across the space of $\theta$ allows one to then capture the uncertainty of the model given a query-document pair~\cite{BayesianGeneralNNWilson,BayesianMCDropoutGal,BayesianUncertaintyExampleaMitros,BayesianUncertaintyExamplebKendall}.

We therefore propose using a Bayesian approach to retain this uncertainty over our model by modeling the full posterior, $P(\theta|\dcmc{D})$ with
\begin{equation}
P(\theta|\dcmc{D}) 
= 
\frac{P(\dcmc{D}|\theta)P(\theta)}{P(\dcmc{D})} 
= 
\frac{P(\dcmc{D}|\theta)P(\theta)}{\int_{\theta} P(\dcmc{D}|\theta)P(\theta) d\theta}.
\end{equation}
The advantage of modeling the full posterior is that we are then able to consider different parameterizations of the model weighted by how well the data supports such a weight configuration. At prediction, our posterior allows us to account for the likelihood of each paramaterization of our retrieval model through the marginalized predictive distribution:
\begin{equation}
P(y|x,\dcmc{D}) = \int_{\theta} P(y|x,\theta) P(\theta|\dcmc{D}) d \theta.
\end{equation}
We then exploit this to capture the retrieval model's uncertainty at retrieval time.

Computing the posterior, specifically $\int_{\theta} P(\dcmc{D}|\theta)P(\theta) d\theta$, in neural architectures is intractable, and so we use an approximation scheme $q(\theta) \approx P(\theta|\dcmc{D})$, called variational inference~\cite{BayesianVI}. The objective of variational inference is to find some $q(\theta)$ that sufficiently fits the data while minimizing the KL divergence to the true posterior $P(\theta|\dcmc{D})$ through the evidence lower bound (ELBO),
\begin{equation}
\text{ELBO}(q) = \mathbb{E}[\log P(\dcmc{D}|\theta)] - KL(q(\theta)||P(\theta)).
\label{eq:ELBO}
\end{equation}

\subsubsection{Monte Carlo Dropout} \label{sec:mcdropout}

We approximate the posterior distribution via Monte Carlo (MC) sampling based on dropout (MC-Dropout) which is a stochastic regularization technique~\cite{BayesianMCDropoutGal}. We then leverage a recent result by~\citet{BayesianLastLayerKristiadi} which bounds the confidence of the model in ReLU networks~\cite{core_relu_hinton10} while simultaneously reducing the computation cost of conventional Bayesian uncertainty estimation.

MC-Dropout approximates the posterior by inducing a distribution for each weight as a mixture of two simpler distributions. Letting $\theta = {\mathbf{W_i}}_1^L,~ \mathbf{W}_i \in \mathbb{R}^{K_i-1 \times K_i}$ for an $L$ layer neural architecture, MC-Dropout  models the variational distribution $q$ via
\begin{align}
&o_{i,j} \sim \text{Bernoulli}(p_i) \text{ for } i=1,\ldots,L, j= 1,\ldots K_i\\
&\mathbf{W}_i = \dcmat{M}_i \cdot diag([o_{i,j}])_{j=1}^{K_i}.
\end{align}Here, $o_{i,j}$ are Bernoulli random variables governed by $p_i$, and $\dcmat{M}_i$ are variational parameters to be optimized. $\dcmat{M}$ can be thought of as the mean of $\dcmat{\theta}$ under a Gaussian with almost 0 variance (essentially delta peaks). The combination of the Bernoulli dropout creates a Gaussian distribution with non-zero variance to form on $\mathbf{W}$ and allows us to model a non trivial approximate posterior probability. We further adopt a concrete perspective on MC-Dropout~\cite{BayesianMCDropoutConcreteGal}, and include $\dcvec{p} = {p_i}_1^L$ as another variational parameter to allow the dropout rate to be a learnable parameter.

At this point, Monte Carlo sampling is used to approximate Equation~\ref{eq:ELBO} to get an unbiased estimate over $N$ draws of $\theta$ from our variational distribution:
\begin{equation}
 \int_{\theta} p(y|x,\theta,f)  q(\theta) d \theta \approx \frac{1}{N} \sum_{t=1}^{N}  p(y|x,\hat{\theta}_t, f) .
\end{equation}The real strength of MC-dropout is the minimal change to standard training procedures. If we assume a standard neural network loss for regression and compare it to the ELBO used for variational inference, we observe a close parallel to standard MLE training with regularization~\cite{TishbyEuclidGaussInterp89}:
\begin{equation}
\frac{1}{2}||y - f_\theta (x) ||^2 + ||\theta||^2 \approx -\frac{1}{\tau} p(y|f_\theta, x) + \KL(q(\theta) || p(\theta)).
\label{eq:tishby}
\end{equation}

As the variational parameters are a delta distribution with the mean at $\theta$, we have the property that $W_{MLE} = \dcmat{M}$ as long as we constrain $L_2$ regularization to equal the $\KL$ divergence. Therefore training a standard neural network with dropout is often equivalent to performing variational inference over the weights of the retrieval model. This feature has been used with success in a variety of computer vision tasks to capture model uncertainty, but the unique nature of pairwise loss functions used in IR, specifically the popular pairwise hinge loss, prevents the direct use of MC-dropout for ranking.

In order to apply MC-Dropout without violating Equation~\ref{eq:tishby}, we relax the conventional pairwise hinge loss to facilitate a Gaussian interpretation. In the case of IR, we can view pairwise hinge loss as minimizing the Euclidean distance between the regression goal and a random point within our collection $\dcmc{D}$ conditioned on some initial ranking $r$ (BM25 sampling, random, kNN, etc): 
\begin{align}
\mathbb{E}[||1-f_\theta(x_+) + f_\theta(x_-) ||^2] = \mathbb{E}[||y-f_\theta(x_+) + f_\theta(X) ||^2]\\ \st~X\sim p(\dcmc{D}|r) .
\end{align}

If we fix $X=x_-$ and define a new function $g_\theta(x_+,x_-) = $ $f_\theta(x_+) + f_\theta(x_-)$, we can treat pairwise hinge loss as a standard regression optimization over $g$. This allows us to still optimize the negative log likelihood of a Gaussian, satisfying the distribution requirements for MC-Dropout:
\begin{equation}
\frac{1}{2}||y - g_\theta (x_+, x_-) ||^2 = -\frac{1}{\tau} p(y|f_\theta, x_-,x_+) + c.
\label{eq:GaussianHinge}
\end{equation}
At evaluation time, setting $g_\theta(x,0)$ where we have a deterministic output of $0$ for $x_-$, we can uncover the uncertainty of $f_\theta(x_+)$. We also consider pairwise cross entropy loss where the direct parallel to its probabilistic interpretation allows for a standard application of MC-Dropout.

\subsection{Efficient Ranking Uncertainty}

As retrieval models have become increasingly computationally demanding~\cite{BERT,repbertZhan,pretrainedBertLin}, the multiple passes required for MC-dropout are not always feasible. In order to retain the effective uncertainty information, we leverage a recent theorem from~\citet{BayesianLastLayerKristiadi} which demonstrates that in the case of binary classification, only the last layer of a model needs to be Bayesian to capture uncertainty information and correct overconfidence. When this occurs, as the test data becomes increasingly distant from the well fit training data, the estimates approach a distribution determined only by the mean and largest eigenvalue of $\mathbf{\theta}$:
\begin{thm}
Let $g: \dcbb{R}^d \rightarrow \dcbb{R}$ be a binary linear classifier defined by $g(\phi(x)) :=  \dctrans{\theta} \phi(x)$, where $\phi: \dcbb{R}^{n}\rightarrow \dcbb{R}^d$ is a fixed ReLU network and let $\dcmc{N}(\theta|\dcvec{\mu},\dcmat{\Sigma})$ be the Gaussian approximation over the last layer's weights with eigenvalues of $\dcmat{\Sigma}$ as $\lambda_1 \leq \ldots \leq \lambda_r$. Then for any input $\dcvec{x} \in \dcbb{R}^n$and $\delta > 0$,
$$
\lim_{\delta \rightarrow \infty} \sigma(|z(\delta \dcvec{x})|) 
\leq \sigma \Big{(} \frac{||\dcmat{\theta}_\mu||}{\sqrt{\pi/8 \lambda_1}} \Big{)}
$$

\label{thm:relu}
\end{thm} 

 While the theoretical result is proven only for the binary classification case, the authors demonstrate its application to softmax with multiple classes. In this work, we demonstrate its empirical validity to the case of ranking with pointwise evaluation where we are able to capture uncertainty information with minimal computation cost.

\subsection{Ranked List Uncertainty Calibration}

Having introduced an approximate Bayesian retrieval framework using only two dropout layers, the usefulness of this uncertainty information partially depends on how well \textit{calibrated} these document score estimates are with respect to the actual accuracy~\cite{CalibrationNNGuo}. Referred to as expected calibration error (ECE), a neural model's calibration is often modeled via binning estimates from [0,1] into $M$ equally distributed buckets, $B_m$, and measuring how much the model's confidence in this estimate deviates from the accuracy. I.e. all predictions with a confidence of $p = 0.3$ should have a mean accuracy of $30\%$. This evaluation is represented as
\begin{equation}
ECE = \sum_{m=1}^M \frac{|B_{m}|}{n} 
\Bigg{|}\frac{1}{B_m} \sum_{i\in B_m} \mathbf{1}(\hat{y}_i = y_i) - 
\frac{1}{B_m} \sum_{i\in B_m} \hat{p_i} 
\Bigg{|}
\label{eq:ece}
\end{equation}
for $n$ samples. However, ECE does not effectively measure a ranking model's calibration for a number of reasons. Neural relevance scores are not calibrated across queries, so document relevance is not distributed on a scale from 0 to 1 where each interval within this range is just as important. This is partially why pairwise or listwise training is substantially more effective than pointwise relevance classification as it is the relative comparison across documents that is most effective, which ECE does not capture. A possible remedy to this issue is to take the softmax over all document scores to force a distribution in $[0,1]$. However, this is again inconsistent as one can reduce the confidence in any individual document by increasing the total number of documents when taking the softmax. We therefore model uncertainty in a pairwise fashion, where calibration is measured between scored documents from the same query via our proposed expected ranking calibration error (ERCE): 
\begin{equation}
\begin{split}
ERCE = \sum_{m=1}^M \frac{|B_{m}|}{n} \Bigg{|} \frac{1}{|B_m|} & \sum_{(D_i,D_j) \in B_m} P(D_i > D_j)\\ &- \frac{1}{|B_m|} \sum_{(D_i,D_j) \in B_m} \mathbf{1}(D_i > D_j)  \Bigg{|}.
\end{split}
\label{eq:erce}
\end{equation}
This allows for a consistent calibration error while still accounting for relevance score distributions being conditioned on queries. The indicator function is defined as
\begin{equation*}
\mathbf{1}(D_i > D_j) = 
\begin{cases}
1 & \text{if ranking}~D_i~\text{above }~D_j~\text{ increases MAP} \\
0 & \text{if ranking}~D_i~\text{above }~D_j~\text{ decreases MAP}, \\
\end{cases}
\end{equation*}
where MAP is mean average precision. This formulation removes all comparisons between pairs of relevant documents, pairs of non-relevant documents, and documents scored from different queries, which allows for measuring only the calibration between relevant and non-relevant pairs conditioned in the same query. In the case of deterministic models which do not have a probabilistic perspective on relevance, we use the pairwise softmax function to calculate $P(D_i > D_j)$.

\subsection{Risk Aware Rerankings}

As each document now has a predictive distribution, we are able to rerank a set of candidates based on a user defined allotted risk. One difficulty of this task is that variance can substantially differ across queries which leaves a linear combination of the type $\lambda\mu+(1-\lambda)\sigma^2,~\lambda \in [0,1]$ ill-suited for robust probabilistic rankings. In order to normalize across all query types and outputs, we approach this problem using the cumulative distribution function (CDF) over scores $s$, $F_S : \dcmc{R} \rightarrow [0,1]$, which maps a score $s$ to the probability of the document achieving a score less than or equal to $s$, i.e., $F_{S}(s) = P(S \leq s)$. This representation has the advantage of normalizing scales to a range of $[0,1]$ across multiple queries and facilitates query agnostic measures of uncertainty~\cite{EvalRLJordan20}. We then rank candidate documents using Conditional Value at Risk (CVaR), 
$$\text{CVaR}_\alpha (S) = \mathbb{E}[S | S \geq F^{-1}_{S}(\alpha)],$$
where $F^{-1}_{S}$ is the inverse CDF of sampled scores. This takes the expected score of a document from the top or bottom $(1-\alpha)\%$ of samples for an optimistic or pessimistic view of outcomes and is often used for risk aware planning or decision making~\cite{cvarExampleYinlam15,cvarRockafellar00}. As CVaR is a coherent risk measure, it satisfies monotonicity and sub-additivity properties and allows us to extend individual document risk to bound total risk for the entire ranked list: $$\text{CVaR}_\alpha(S_1, \ldots, S_n) \leq \text{CVaR}_\alpha(S_1) + \ldots + \text{CVaR}_\alpha(S_n).$$

\section{Experiments}

We examine four attributes with respect to efficient Bayesian retrieval. First, we examine the hypothesis that the mean weight of the dropout model is equivalent to its deterministic variant. Second, we study the ranking calibration error in the same manner. Third, we investigate the impact of CVaR$_\alpha$ in both optimistic and pessimistic settings for risk-aware rankings. And finally, we evaluate the usefulness of this uncertainty information under the downstream task of cutoff prediction~\cite{choppycutoffBahri20,bicut_cutoffLien19}.

\subsection{Data}
\begin{table}[]
\caption{Training, validation, and test statistics for the collections used. No training, fine-tuning, or validation is performed for any retrieval models on Robust04 thus use no validation set is used.}
\begin{tabular}{lrrr}
\toprule
 \textbf{Collection}& \textbf{Documents} &  \textbf{Validation}  & \textbf{Test}  \\ 
 \hline
 MS~MARCO   & 9M &  6,980 &  48,598  \\
 TREC 2019 DLT &  9M  & 6,980  & 43 \\
 Robust04 & 0.5M  & - & 250 \\
 \bottomrule
\end{tabular}

\label{tab:collection}
\end{table}

We utilize three collections for our experiments. Evaluating retrieval and cutoff performance, we use the MS~MARCO~\cite{MSMARCOdataset16} passage dataset, the TREC 2019 Deep Learning Track (DLT)~\citep{trec2019overview} based on the original MS~MARCO dataset, and Robust04 with its corresponding title queries. While MS~MARCO is commonly used to evaluate performance of retrieval models, the one hot relevance judgements limit the investigation of uncertainty in ranking. We therefore use this dataset as a general training collection and use both the TREC 2019 Deep Learning Track and Robust04 collections with their more fine grained relevance judgements to examine uncertainty properties with risk-aware re-rankings, calibration analysis, and cutoff prediction.

With respect to training, validation, and test splits, in the case of MS~MARCO, we evaluate with the same validation and test splits as ~\citet{hofstatter2019effect}. We do not fine tune or validate on TREC 2019 DLT nor Robust04, and therefore all queries are used to test the models. As demonstrated by Yilmaz et al.~\cite{CrossDomainBERTIRLin2019}, an MS~MARCO trained BERT architecture is an effective retrieval model for Robust04 which results in a non-trivial evaluation. We transform each full document into the first 512 tokens as its representative passage in a similar fashion to~\citet{BERTRobustFirstP}. Statistics of the collections are provided in Table~\ref{tab:collection}.

\subsection{Models}

We examine two representative retrieval architectures using our uncertainty framework: BERT~\cite{BERTPassageRankCho} and Conv-KNRM~\cite{convKNRMDai18}. BERT represents the current trend towards large pre-trained transformer architectures, and Conv-KNRM provides insight into how well last layer MC-Dropout works on retrieval models which rely on hand crafted similarity functions as input into the Bayesian layer. Given hardware constraints, we evaluate on tiny and mini BERT (BERT-L2, BERT-L4) versions which achieve close to full BERT performance~\cite{BERTwellReadStudentsTinyMiniTurc2019}.

\subsection{Baselines}

For each Bayesian model, we evaluate with respect to the deterministic version of the architecture denoted by subscript ${D}$. While the work by \citet{RiskAwareIRZhu09} discusses the concept of risk aware reranking by the variance of a language model, this view does not directly apply to the case of modern deep retrieval architectures. In the case cutoff prediction, we include a modern interpretation of their work with a generative BERT-to-Transformer architecture to provide context

For the application task of cutoff prediction, we use the architecture proposed by \citet{liu2019text}, where the document is encoded by the mini/tiny BERT and the query is decoded term-by-term through a 4-layer Transformer decoder conditioned on the encoded document. We train the model following the same procedure as in \citet{nogueira2019document}, and similar to \citet{izacard2020leveraging} consider relevance as $\log P(Q|D)$, namely the sum of the logarithms of the next term generation probabilities. An uncertainty interpretation for these generative architectures produces entropy values over $P(Q|D)$, and provides useful information as we can compare self reported aleatoric uncertainty to both aleatoric and epistemic uncertainty from our Bayesian approximation.

\subsection{Evaluation}

\subsubsection{Retrieval and Reranking: }We evaluate each model using mean reciprocal rank (MRR) and normalized discounted cumulative gain (nDCG@n). MRR was selected due to the single judged relevant passage per query in MS~MARCO, while nDCG@20 was used for Robust04 and nDCG@200 for TREC 2019 DLT. We use a large cutoff for TREC to better capture less confident document relevance judgements, and use the traditional lower cutoff for Robust04 as it is already out of distribution.

\subsection{Calibration:} We use an adaptive binning scheme to dynamically create 10 equally filled buckets $B$.

\subsubsection{Cutoff Prediction:} We use a state-of-the-art transformer based prediction model, Choppy, to determine where to cut a ranked list to maximize $F_1$~\cite{choppycutoffBahri20}. As the performance is a function of the oracle, we report $$\frac{F1_{Choppy}}{F1_{Oracle}}$$ where $F1_{Choppy}$ and $F1_{Oracle}$ are the $F_1$ scores produced by the cutoff identified by Choppy and the oracle cutoff, respectively. We report our results using a ranked list of the top 200 documents for each model.  In case of the Bayesian models, each document is represented as $< \mu, \sigma, \delta, H(S) >$,
with $\delta$ as the skew and $H(S)$ as the entropy over the score distribution $S$.

\subsection{Hyperparameters and Training}

For each model architecture, we use the same hyperparameters used for the deterministic variant. BERT models were trained with a learning rate of $1\times 10^{-4}$ while the Conv-KNRM used $1 \times 10^{-3}$ using Adam optimization~\cite{adamKingma15}. We use two additional feed-forward layers after the main model to facilitate last layer MC-Dropout of sizes $[K,K], [K,f]$ where $K$ is the output size of the kernel features in Conv-KNRM or the hidden dimension of BERT, and $f$ is the final output dimension of the architecture. In case of the stochastic models, a single sample was used during training and $150$ samples were used at inference time for each query-document pair.

\section{Results}
\begin{table*}[h]
\caption{Comparison of MRR and nDCG@n performance between deterministic ($_{D}$) and mean Bayesian MC-Dropout ($_B$) model performance. Document scores were calculated as mean performance over 150 samples for the stochastic models. $n$ was 200 for TREC 2019 DLT and 20 for Robust04. $^*$ represents statistical significance of $p<0.05$ using paired t test.}
\centering
\begin{tabular}{lll|ll|llll}
\toprule
 \textbf{Collection}& \multicolumn{6}{c}{\textbf{Models}}  \\ 
 & \multicolumn{6}{c}{\textbf{MRR}} \\
            & BERT-L2$_{D}$ & BERT-L2$_{B}$  & BERT-L4$_{D}$ & BERT-L4$_{B}$ & Conv-KNRM$_{D}$ & Conv-KNRM$_{B}$ \\
            \cmidrule{2-7}
 
 MS~MARCO                                & 0.305 & 0.301~(-1.3\%) & 0.308  & 0.308~(+0.1\%) & 0.279 & 0.280~(+0.4\%) \\
 TREC 2019 DLT                           & 0.912 & 0.916~(+0.5\%) &  0.929 & 0.936~(+0.8\%) & 0.900  & 0.901~(0.0\%)\\
 \hline
 MS~MARCO $\rightarrow$ Robust04         & 0.617 &  0.628$^*$~(+1.8\%) & 0.657 &  0.640$^*$~(-2.6\%)  & 0.591 & 0.598~(+1.2\%)   \\
 \bottomrule
  & \multicolumn{6}{c}{\textbf{nDCG@200,20}} \\
            & BERT-L2$_{D}$ & BERT-L2$_{B}$  & BERT-L4$_{D}$ & BERT-L4$_{B}$ & Conv-KNRM$_{D}$ & Conv-KNRM$_{B}$ \\
            \cmidrule{2-7}
 
 MS~MARCO                                & 0.398 & 0.395 (-0.8\%) & 0.401  & 0.401 (-0.1\%) & 0.380 & 0.377 (-0.8\%) \\
 TREC 2019 DLT                           & 0.582 & 0.576 (-1.0\%) &  0.582 & 0.581 (-0.2\%) & 0.565  & 0.567 (+0.4\%)\\
 \hline
 MS~MARCO $\rightarrow$ Robust04         & 0.431 &  0.433 (+0.7\%) &  0.434 & 0.431 (-0.7\%) & 0.425 & 0.426 (+0.2\%)     \\
 \bottomrule
\end{tabular}

\label{tab:results}
\end{table*}

In this section, we first compare our Bayesian interpretation against baseline deterministic models to ensure the stochastic nature does not significantly degrade actual retrieval performance. We then study the behaviour of uncertainty across relevance scores and evaluate risk aware re-ranking. We further report ERCE measures prior to discussing downstream usefulness with cutoff prediction.

\subsection{Comparison to Deterministic Models} \label{sec:parity}

While we have discussed the many benefits of well calibrated measures of uncertainty, the introduction of stochastic scoring should not harm actual retrieval performance. As such, we address our first hypothesis: \textit{Does efficient MC-Dropout at inference time result in the same mean performance as a deterministic retrieval model?} As Table~\ref{tab:results} demonstrates, mean MC-Dropout document scoring achieves parity with its deterministic variants across collections and architecture types. While there is some discrepancy with BERT-L2$_{B}$ slightly under performing and Conv-KNRM$_{B}$ and BERT-L4$_{B}$ outperforming the deterministic versions, the small differences paired with the nature of stochastic gradient update demonstrates that this framework is a safe way to include uncertainty information into a variety of architectures without the risk of deteriorating performance. This parity also confirms that using MC-Dropout empirically satisfies Equation~\ref{eq:tishby}.

On Robust04, we observe the largest difference in performance of approximately -2.6\% for BERT-L4. This result highlights that uncertainty aware models do not inherently perform better on out-of-distribution data; however, we discuss how the models are able to convey their uncertainty and how to leverage this information to improve performance.

\subsection{Risk-Aware Ranking}

\begin{table}[]
\caption{Risk-aware rerankings of the top 200 candidate documents for each query using CVaR. CVaR$_1$ indicates standard mean score, CVaR$_+$ and CVaR$_-$ represents taking the optimistic and pessimistic perspectives above and below $\alpha$ accordingly. We report nDCG@200 for TREC DLT and nDCG@20 for Robust04. * denotes statistical significance with respect to baselines with $p < 0.05$ using paired t test.}
\centering
\begin{tabular}{lllll}
\toprule
 \textbf{Collection}    & \textbf{Model}         & \textbf{CVaR}$_{+}$ & \textbf{CVaR}$_{-}$ & \textbf{CVaR}$_1$ \\
 
 \multirow{ 6}{*}{TREC 2019 DL}     & BERT-L2$_{D}$   &  0.582 & 0.582 & 0.582  \\
                                    & BERT-L2$_{B}$  &  0.597* & \textbf{0.598}* & 0.576  \\
                                   \cmidrule{2-5}
                                    & BERT-L4$_{D}$   &  0.582 & 0.582 & 0.582  \\
                                    & BERT-L4$_{B}$  &  \textbf{0.606}* & 0.605* & 0.581  \\
                                    \cmidrule{2-5}
                                    & Conv-KNRM$_{D}$ &  0.565 & 0.565 & 0.565  \\
                                    & Conv-KNRM$_{B}$&  0.584* & \textbf{0.585}* & 0.567 \\
                                   \midrule \midrule
     \multirow{ 6}{*}{Robust04}     & BERT-L2$_{D}$   &  0.398 & 0.398 & 0.398  \\
                                    & BERT-L2$_{B}$  &  0.411* & \textbf{0.412}* & 0.402  \\
                                    \cmidrule{2-5}
                                    & BERT-L4$_{D}$   &  0.400 & 0.400 & 0.400  \\
                                    & BERT-L4$_{B}$  &  \textbf{0.407}* & \textbf{0.407}* & 0.395  \\
                                   \cmidrule{2-5}
                                    & Conv-KNRM$_{D}$ &  0.382 & 0.382 & 0.382  \\
                                    & Conv-KNRM$_{B}$&  \textbf{0.404}* & 0.403* & 0.386  \\
 \toprule 
                                                  
 \bottomrule
\end{tabular}

\label{tab:cvar}
\end{table}

Incorporating the CVaR based ranking (Table~\ref{tab:cvar}) allows us to leverage the model's reported uncertainty, consistently improving performance across all collections. After accounting for the variance and skew of each document's score distribution, MC-Dropout based models significantly improve performance both with respect to their mean performance, but also their deterministic baseline models by 3-5\%. This performance increase is present in both the optimistic (CVaR$_+$) and pessimistic (CVaR$_-$) setup, suggesting that focusing on the tails of either end of the distributions provides pertinent uncertainty information with respect to the model outputs. Interestingly, we see a similar change in performance for the Robust04 collection, suggesting that MC-Dropout models are equally capable of expressing risk on the data used to train the model in addition to collections where all documents are out of distribution.

\begin{figure*}
    \centering
    \includegraphics[width=\textwidth]{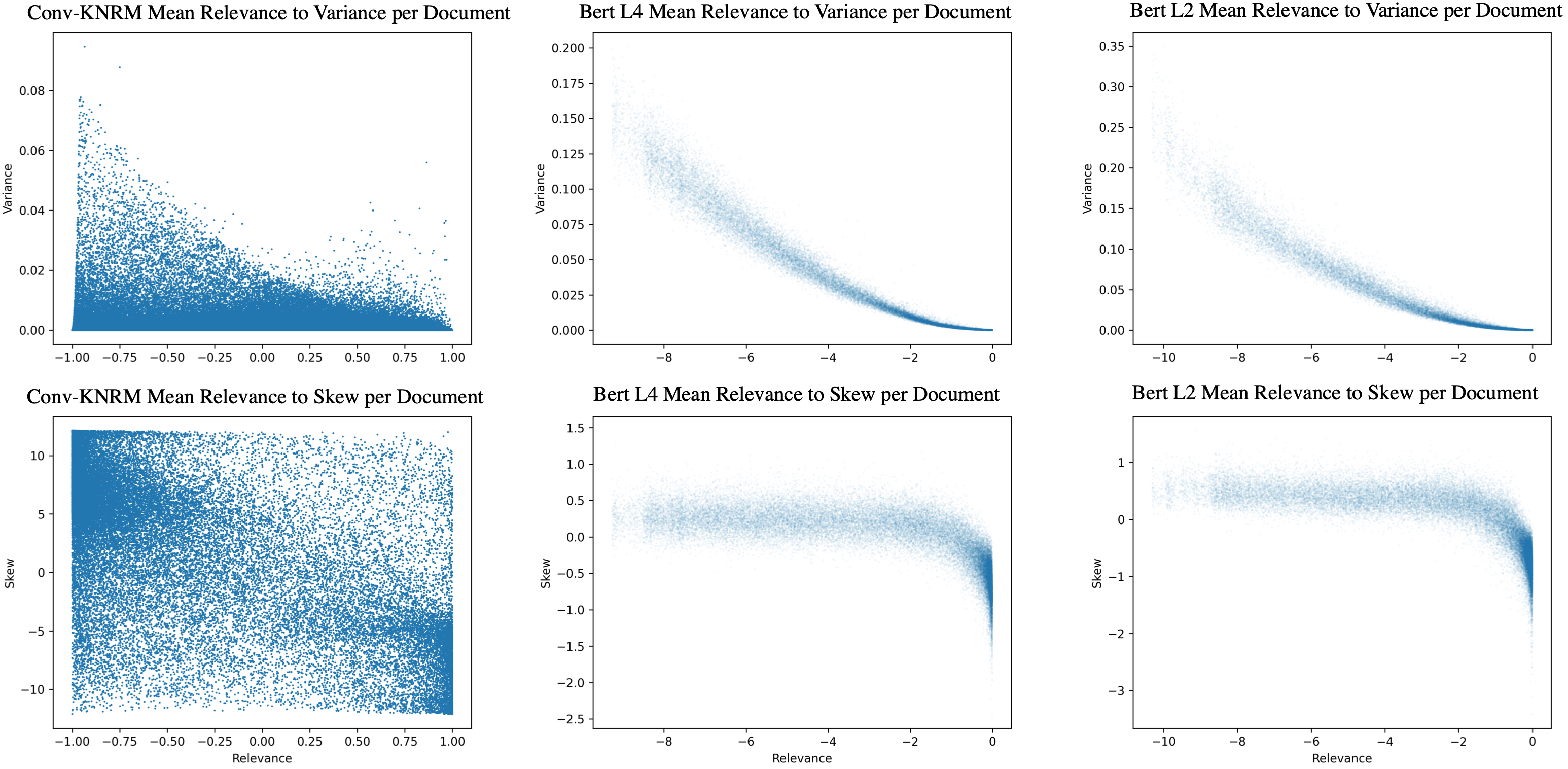}
    \caption{The mean to variance and mean to skew relationship for each document scored by Conv-KNRM$_{B}$ (top), BERT-L2$_{B}$ (middle), and BERT-L4$_{B}$ on TREC 2019 DLT.}
    \label{fig:var}
\end{figure*}

\begin{figure}
    \centering
    \includegraphics[width=.72\linewidth]{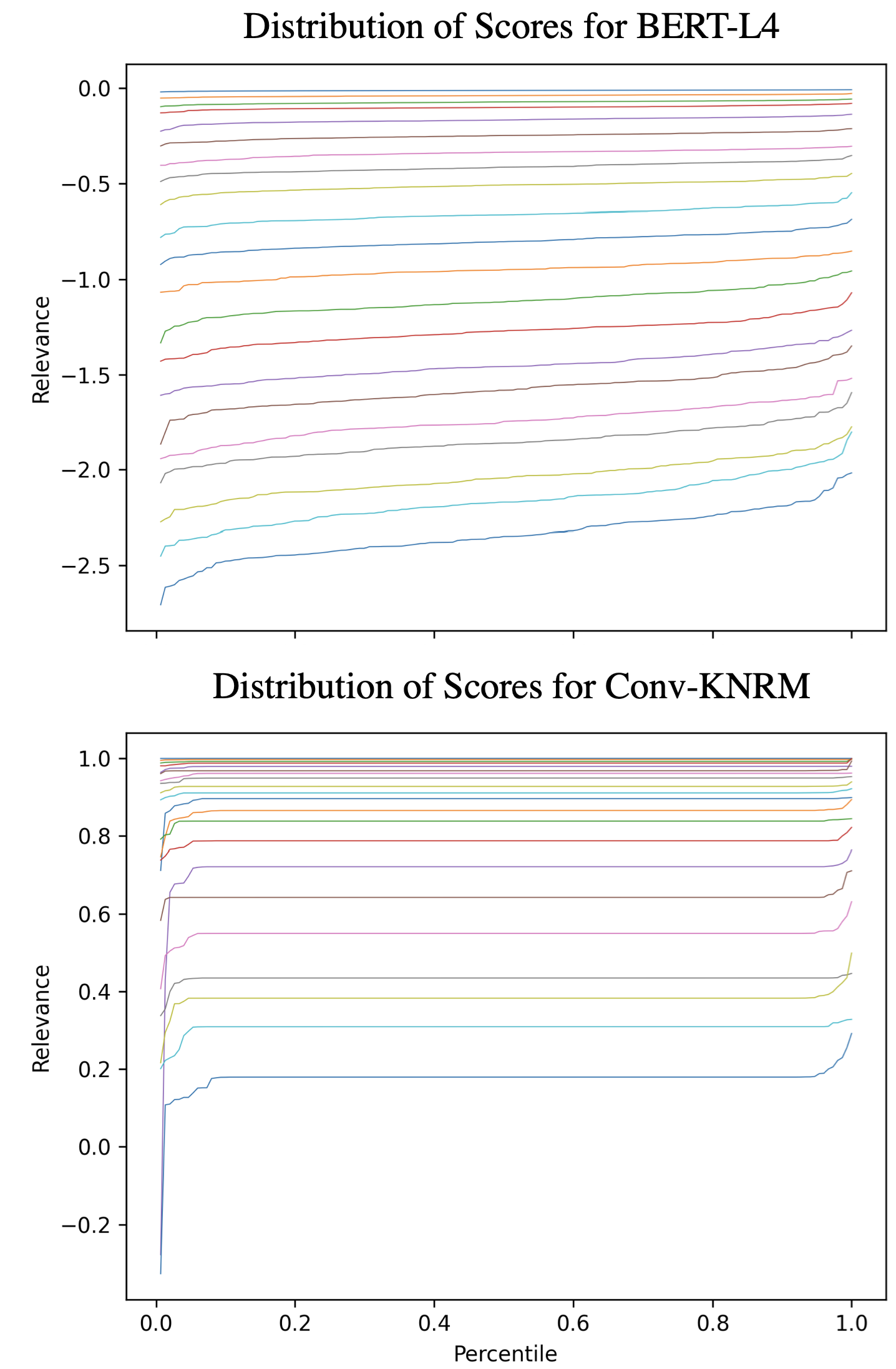}
    \caption{Empirical CDFs of a subset of documents for a single query on  BERT-L4$_{B}$ (top),  Conv-KNRM$_{B}$ (bottom) for the TREC 2019 DLT dataset. Documents were selected for every 10$^\text{th}$ rank position $\mathbf{(0,10,20,30\ldots, 200)}$.}
    \label{fig:cdf}
\end{figure}

To provide additional insight into how this risk aware re-ranking functions over different candidate ranking positions, we plot the relationship between $\mu$ to $\sigma^2$ and $\mu$ to skew $\delta$ in Figure~\ref{fig:var}. It is in these figures that we notice the impact that the neural architecture has on the uncertainty of the documents. Both BERTs have a direct relationship with predicted relevance such that the model is most certain about highly relevant documents, and a non-linear increase in uncertainty as documents move further away from the query. With this increase we also see the scores converge to a normal distribution. The top most relevant documents are highly skewed with low variance within the range of 0 to -0.1 relevance score with the majority of its mass on the right of the distribution as indicated in Figure~\ref{fig:cdf}. As the variance increases and relevance scores drop, the score distribution follows Theorem~\ref{thm:relu} and approaches a normal distribution.

With respect to Conv-KNRM$_{B}$, the same pattern is present but not as salient. The upper bound of variance continues to grow while a large portion of non-relevant documents still have very low relevance. However, the asymmetric nature across $tanh$ used in ConvKNRM  demonstrates that uncertainty is still being expressed over both the handmade kernel features and the condensing nature of $tanh$. This same asymmetry is found in the skew plot, with a greater number of documents expressing high positive skew than the fewer highly relevant documents with large negative skew. We hypothesize that the polarizing skew values is due to the high gradient of $tanh$.

The variance and skewness trends are reinforced as the risk-aware CVaR consistently increases metric performance as we increase the $n$ cutoff for nDCG@n, suggesting that risk based re-ranking can be most utilized for high recall tasks or where effective performance is required outside of the top few documents. This result is unsurprising when considering the inverse CDF plots in Figure~\ref{fig:cdf}. The highly ranked documents have an almost point distribution while the lower ranked documents exhibit significant uncertainty.

\subsection{Calibration}

\begin{table*}[]
\caption{Expected ranking calibration error (ERCE) for BERT models with respect to their deterministic variants (lower is better).}
\centering
\begin{tabular}{lll|ll|llll}
\toprule
 \textbf{Collection}& \multicolumn{6}{c}{\textbf{Models}}  \\ 
            & BERT-L2$_{D}$ & BERT-L2$_{B}$  & BERT-L4$_{D}$ & BERT-L4$_{B}$ & Conv-KNRM$_{D}$ & Conv-KNRM$_{B}$ \\
            \cmidrule{2-7}
 
 TREC 2019 DLT                           & 0.703 & \textbf{0.465} &  0.700 & \textbf{0.507} & 0.519  & \textbf{0.452}\\

 MS~MARCO $\rightarrow$ Robust04         & 0.493 &  \textbf{0.396} &  0.477 & \textbf{0.395} & \textbf{0.256} & 0.264   \\
 \bottomrule
\end{tabular}

\label{tab:calibration}

\end{table*}

We now inspect the expressiveness of a stochastic retrieval model's uncertainty to answer \textit{Does efficient MC-Dropout improve uncertainty calibration for IR models?} We record ERCE in Table~\ref{tab:calibration} for all model and collection permutations and observe a substantial decrease in calibration error resulting in approximate Bayesian models being \~30\% more calibrated. This confirms our hypothesis that Bayesian retrieval models will have better expressiveness of their confidence. It follows from recent results in computer vision~\cite{CalibrationNNGuo}, where non Bayesian models are poorly calibrated. These results explain the natural separation of mean document scores in Figure~\ref{fig:cdf} as the highly relevant documents are clustered together while the lower ranked documents show a greater spread. One perspective on this is that the increased spread of scores across the range of relevance expressed by the model can be viewed as the relative confidence that they are ranked in the right order.

\begin{table*}[ht!]
\caption{Choppy~\cite{choppycutoffBahri20} performance as a percentage of oracle cutoff under the F1 metric. $G$ is the generative baseline using BERT-L2 and BERT-L4, and $^*$ denotes $p<.05$ significance using t test with respect to baseline variants of the same architecture.}
\centering
\begin{tabular}{llll|lll|llll}
\toprule
 \textbf{Collection}& \multicolumn{8}{c}{\textbf{Models}}  \\ 
            & BERT-L2$_{D}$ & BERT-L2$_{B}$ & BERT-L2$_{G}$  & BERT-L4$_{D}$ & BERT-L4$_{B}$  & BERT-L4$_{G}$ & Conv-KNRM$_{D}$ & Conv-KNRM$_{B}$ \\
            \cmidrule{2-9}
 
TREC 2019 DLT                            & 77.3\% & \textbf{80.6\%}$^*$ & 74.5\% & 73.6\% & \textbf{79.9\%}$^*$ & 73.8\% & 75.3\%  & \textbf{86.4\%}$^*$\\

Robust04                                 & 74.1\% &  \textbf{78.9\%}$^*$ & 62.4\% &  75.4\% & \textbf{78.1\%}$^*$ & 63.3\% & 66.2\%  & \textbf{77.7\%}$^*$  \\
 \bottomrule

\end{tabular}

\label{tab:cutoff}

\end{table*}

\subsection{Downstream Application: Cutoff Prediction}

Having discussed the improved calibration and risk based re-ranking, we now address our question \textit{Is uncertainty information  actionable in the context of downstream tasks?} To do so, we use the cutoff prediction task where the objective is to find a cutoff point in a ranked list that maximizes some non-monotonic metric. The motivation, discussed by~\citet{bicut_cutoffLien19}, is that at some point, a neural model loses effectiveness as documents move further away from the query. Using the Choppy cutoff predictor from~\citet{choppycutoffBahri20}, we compare the cutoff performance between the information contained in a deterministic model, a modern version of \citet{RiskAwareIRZhu09}'s risk based language model, and our Bayesian model's score distribution in Table~\ref{tab:cutoff}. As shown, we observe a significant improvement to the upper performance bound (oracle) under the Bayesian framework when compared to the deterministic models. As indicated in Section~\ref{sec:mcdropout}, the mean weights of the MC-Dropout models closely approximates those of their deterministic cousins. This suggests, combined with the visual inspection of Figure~\ref{fig:bayesian}, that the key determining factors are the variance, entropy, and skew values as a function of document relevance.

Examining the cutoff performance across TREC DLT 2019, we note an approximate 9\% increase in cutoff accuracy when including the additional uncertainty information, confirming our hypothesis that the uncertainty information displayed by the MC-Dropout models can be used in downstream decision making. Moving to the Robust04 results, we see a greater increase in comparative performance. As the retrieval models are now out of distribution, there exists significant variance across all ranking positions which introduces noise into the additional dimensions. Following the same trends discussed in risk-aware re-ranking, we see the greatest improvement using an initial ranked list of the top 200 documents. As we decrease the set of candidate documents to be re-ranked, the performance difference between deterministic and Bayesian variants closes. At the top 50 candidates we record only a 2\% difference in performance across model frameworks.

Remarking on the related work of capturing uncertainty through generative retrieval models, we further highlight the performance gap between the generative BERT and the Bayesian BERT models~\cite{RiskAwareIRZhu09,izacard2020leveraging}. Using the generative framework introduced by~\citet{liu2019text} where the BERT component uses BERT-L2 or BERT-L4, its scores and entropy values representing uncertainty over relevance fail to achieve the same level of calibrated uncertainty as the approximate Bayesian approach. This highlights the finding by \citet{GenKnowWhatTheyDontKnowNalisnick19} that generative models, while capable of expressing uncertainty, are often overconfident in their own self-estimates, resulting in uncertainty measures which are not as robust when compared to those made by Bayesian models. Further, the uncertainty values self-reported by the model are substantially worse when out of distribution on Robust04 with close to a 15\% degradation in cutoff performance, which demonstrates the robust uncertainty present in the Bayesian distributions.

\subsection{Efficiency}

As one of the primary contributions of this work is the efficient modeling of uncertainty, one of the most significant obstacles to be addressed is the computational cost of scoring each query-document pair $n$ times. We benchmark the additional compute cost for our last-layer MC-Dropout on a GTX 1080ti. While not completely free, the additional cost of running 100 additional samples is $0.326\pm 0.012~\mu s$ for Conv-KNRM4$_{B}$ and $0.368\pm 0.016~\mu s$ for BERT-L4$_{B}$, L2$_{B}$, or for any other large transformer architecture as the additional cost is a function of the final output dimension, not of the retrieval model itself.

\section{Conclusion}
In this paper, we introduced an efficient Bayesian framework to estimate epistemic and aleatoric uncertainty in retrieval models. We demonstrated that query-document uncertainty can be modeled using only the last two layers of a neural model, allowing for its use in state-of-the-art retrieval models relying on BERT -- be it pre-trained or as part of the actual retrieval architecture. The performance of these stochastic models stays reasonably close to their deterministic versions while offering substantially more information per document score. Furthermore, the actual scores themselves are better calibrated with each other allowing for a more accurate comparison between documents. These properties enable improved performance on ranking via risk-aware reranking in addition to the downstream task of cutoff prediction when compared to the deterministic versions.

As this approximate Bayesian inference is efficient and conveys useful information for both fully distributed and handcrafted models, there exists a promising body of future work incorporating stochastic models for fairness~\cite{FairnessEvalLipani16, StochRankingsDiaz20}, diversity~\cite{DiversityEvalIRClarke08}, transparent search~\cite{TransparencyIREickhoff20}, dialogue agents~\cite{ConversationalIRHashemi20} and improved sample efficiency when training neural retrieval models~\cite{ANCExiong2020}. Lastly, we hope to explore the impact of uncertainty modeling in situations where the retrieval model acts as an information gathering agent in larger systems.

\section{Acknowledgements}
This research is supported in part by the NSF (IIS-1956221), ODNI and IARPA via the BETTER program (2019-19051600004). The views and conclusions contained herein are those of the authors and should not be interpreted as necessarily representing the official policies, either expressed or implied, of NSF, ODNI, IARPA or the U.S. Government.

\FloatBarrier
\bibliographystyle{ACM-Reference-Format}
\bibliography{main}

\end{document}